\documentclass[fleqn,usenatbib]{mnras}

\usepackage[normalem]{ulem}

\usepackage[T1]{fontenc}

\DeclareRobustCommand{\VAN}[3]{#2}
\let\VANthebibliography\thebibliography
\def\thebibliography{\DeclareRobustCommand{\VAN}[3]{##3}\VANthebibliography}

\usepackage{graphicx}	
\usepackage{amsmath}	
\usepackage{amssymb}	
\usepackage{newtxtext,newtxmath}

\title[Micro-tidal disruption events]{Partial tidal disruption events by stellar mass black holes: gravitational instability of stream and impact from remnant core}

\author[Y-H. Wang et al.]{
Yi-Han Wang$^{1}$\thanks{E-mail: yihan.wang.1@stonybrook.edu},
Rosalba Perna$^{1,2}$,
Philip J. Armitage$^{1,2}$
\\
$^{1}$Department of Physics and Astronomy, Stony Brook
  University, Stony Brook, NY 11794-3800, USA\\
$^{2}$Center for Computational Astrophysics, Flatiron Institute, 162 Fifth Avenue, New York, NY 10010, USA
}

\date{Accepted XXX. Received YYY; in original form ZZZ}

\pubyear{2015}

\begin{document}
\label{firstpage}
\pagerange{\pageref{firstpage}--\pageref{lastpage}}
\maketitle

\begin{abstract}
In dense star clusters, such as globular  and open clusters, dynamical interactions between stars and black holes (BHs) can be extremely frequent, leading to various astrophysical transients. Close encounters between a star and a stellar mass BH make it possible for the star to be tidally disrupted by the BH. Due to the relative low mass of the BH and the small cross section of  the tidal disruption event (TDE) for cases with high penetration, disruptions caused by close encounters are usually partial disruptions. The existence of the remnant stellar core and its non-negligible mass compared to the stellar mass BH alters the accretion process significantly. We study this problem with SPH simulations using the code  {\tt Phantom}, with the inclusion of radiation pressure, which is important for small mass BHs. Additionally, we develop a new, more general method of computing the fallback rate which does not rely on any approximation. Our study shows that the powerlaw slope of the fallback rate has a strong dependence on the mass of the BH in the stellar mass BH regime. Furthermore, in this regime, self-gravity of the fallback stream and local instabilities become more significant, and cause the disrupted material to collapse into small clumps before returning to the BH. This results in an abrupt increase of the fallback rate, which can significantly deviate from a powerlaw. Our results will help in the identification of TDEs by stellar mass BHs in dense clusters.
\end{abstract}




\section{Introduction}
The tidal disruption of stars by  supermassive black holes (SMBHs) in galactic centres has been extensively studied over the past few decades both numerically 
and observationally \citep{komossa15,roth2020,dai2021}.
The properties of the resulting flares from tidal disruption events (TDEs) have been used as probes to study the SMBHs in quiescent galactic nuclei and their surrounding stellar population.

The first direct detection of a binary black hole (BH) merger, by the LIGO observatory \citep{Abbott2016},
has further increased interest in stellar dynamical processes in dense star clusters.
In high-density star clusters, such as young massive clusters and globular clusters, the presence of small-$N$ bound systems increases the rate of close flybys between compact objects (black holes, neutron stars and white dwarfs) and stars \citep{Bacon1996,Fregeau2004}. If the flyby is sufficiently close, the star can be tidally disrupted by the compact object. \citet{Perets2016} suggested that the disruption of stars by stellar mass BHs (described as micro-TDEs) may give rise to long X-ray/gamma ray flares, which could possibly resemble ultra-long gamma-ray bursts. Dynamical studies (e.g. \citealt{Fragione2019,Samsing2019,Fragione2020}) have suggested that stellar TDEs could be important probes of the BH population in star clusters.

The characteristics of the light curves from TDEs are largely determined by the rate at which debris falls back, circularizes, and subsequently accretes onto the BH \citep{Lodato2011,Roth2016}.  
Calculations of the fallback rate often rely on the ``frozen-in"  approximation, according to which the star is destroyed impulsively at the tidal radius, with the debris then following ballistic trajectories \citep{Rees1988}. The accretion rate will track the fallback rate if, additionally: {(i)} the kinetic energy of the returning debris is dissipated efficiently, {(ii)} the material rapidly circularizes, and {(iii)} the viscous timescale in the newly formed disc is short compared to the fallback time.
If the star is fully tidally disrupted by the compact object, then, within these approximations, the gas parcels comprising the disrupted star follow pure Keplerian orbits. During the late stages of the accretion process \citep{Cannizzo1990, Mockler2019,Miles2020}, the fallback rate from full disruptions will asymptote to $t^{-5/3}$ \citep{Phinney1989}.

Hydrodynamical simulations performed by \citet{Guillochon2013} showed that there is a critical penetration factor $\beta=r_{\rm t}/r_{\rm p}$ that separates TDEs into full and partial disruptions, where $r_{\rm t}$ and $r_{\rm p}$ are the tidal radius and pericenter distance of the star, respectively. In partial TDEs, a stellar remnant core survives from the first encounter with the compact object. The surviving remnant stellar core can then interact gravitationally with the returning debris. Due to this additional interaction, the material in the stream no longer obeys the Keplerian trajectory as it falls back to the pericenter. Therefore, the fallback rate from such partial TDEs could be significantly steeper than the $t^{-5/3}$ scaling. Due to the relatively small cross section of high-$\beta$ encounters between stars and compact objects in dense clusters, most of the TDEs by stellar mass compact objects are partial TDEs. This makes the study of partial TDEs by stellar mass compact objects especially  interesting. 

\citet{Coughlin2019} developed a model to analytically calculate the asymptotic temporal scaling of the late time fallback rate of partial TDEs. By solving the equation of motion of the elements in the stream together with the mass conservation law, they concluded that the late time fallback rate from a partial TDE asymptotically scales approximately as $t^{-9/4}$, and is effectively independent of the mass of the remnant stellar core which survives the encounter. This scaling is derived based on several approximations: (i) the self-gravity of the debris can be ignored, (ii) the trajectory of the remnant core is parabolic, (iii) the angular momentum of the stream with respect to the compact object is zero, and (iv) the mass of the remnant stellar core is much smaller than the mass of the compact object. \citet{Coughlin2019} validated those assumptions in the SMBH regime and found that the analytical solution fits the numerical results very well. However, for TDEs by stellar mass compact objects, where the mass of the remnant stellar core is more significant compared to the mass of the compact object, these assumptions become more difficult to maintain. For instance, due to the lower mass of the accreting compact object, the self-gravity of the debris in the stream exerts a stronger impact on the local dynamics of the stream, leading to more prominent local gravitational instabilities. The trajectory of the remnant core could be elliptical. Furthermore, as accretion disks around stellar mass BHs are  known to be pressure-dominated in the highly hyper-accreting regime \citep{Narayan2001,DiMatteo2002,Janiuk2004}, radiation pressure can be dominant in the disks formed during TDEs
from stellar-mass BHs.

In order to study the partial TDE problem from stellar-mass BHs 
we perform a series of hydrodynamic simulations with the SPH code {\tt Phantom} \citep{Price2018}. We implement a new radiative equation of state to better capture the shock heating and radiation from the falling-back material, and use a new method to calculate the fallback rate for partial TDEs which does not rely on the frozen-in approximation.  We derive the fallback rate scalings of partial TDEs and explore the dependence on the BH mass, highlighting the significant differences that are caused by the presence of radiation pressure.

Our paper is organized as follows. Section \ref{sec:method} describes the details of the implementation of the new radiative equation of state in   {\tt Phantom}, the new method to calculate the fallback rate of the partial TDEs, and the simulation setup. We present our results in Section~\ref{sec:results} and summarize our conclusions in Section~\ref{sec:summary}.

\section{Equation of state and fallback rate}
\label{sec:method}
In simulations of TDEs  with the SPH code   {\tt Phantom}, to save computing time and stabilize the initial model of the star, 
a polytropic equation of state is typically used,
\begin{equation}
    P=K\rho^\gamma\,,
\end{equation}
where $K$ and $\gamma$ are global constants. With this option\footnote{ It is selected in the code as the ``isothermal" compiling option.}, the temperature remains low even after shock generation, implying that the shock heating is radiated away immediately after the shock. Therefore shock heating cannot be correctly tracked. On the other hand, if the heat from the shock is not efficiently cooled via radiation, then the local temperature increases significantly and radiation pressure can become dominant over gas pressure. Under these conditions, the inclusion of radiation pressure is required in the code.
In the following (Section~\ref{sec_radiative_cooling}), we show that cooling in the flow is in fact not efficient enough, and hence radiation pressure needs to be included; its implementation in   {\tt Phantom} is described in Section~\ref{sec_implementation}.

\subsection{Radiative cooling}
\label{sec_radiative_cooling}
If the accretion disk is optical thin, the approximation of efficient radiative cooling is good if the timescale of radiative cooling is shorter than the dynamical timescale.

Whether heat can be quickly radiated away depends on the local optical depth of the accretion disk; at the leading order approximation, the optical depth near the apocenter can be estimated as \citep{Piran2015,Ryu2020a},
\begin{equation}\label{eq:tau}
    \tau \sim  \frac{\kappa M_\star}{4\pi a_{\rm min}^2}\,,
\end{equation}
where $\kappa=0.34~{\rm cm}^2{\rm g}^{-1}$ is the Thomson opacity, $M_\star$ is the mass of the star, and $a_{\rm min}$ is the semi-major axis of the most bound material,
\begin{equation}
    a_{\rm min} \sim \gamma \frac{r_{\rm t}^2}{2R_\star}\,.
\end{equation}
In the above, $r_{\rm t} = (M_\bullet/M_\star)^{1/3}R_\star$ is the tidal radius  for a Solar type star, $R_\star$  the radius of the star, and $\gamma$ a geometry factor. A 10 $M_\odot$ black hole yields $\tau\sim 2\times10^{9}$ with a geometry factor of 1. However, the outskirts of the accretion disk are more optically thin. To make a more precise estimate, we also run a test simulation of a TDE by a 10~$M_\odot$ BH with the simulation setups described in Section~\ref{sec:setups} and calculate the optical depth of the emission region from the snapshot. The optical depth is calculated via the equation, 
\begin{equation}
    \tau(r) = \int_{r}^{+\infty} n_{e}(r^\prime) \sigma_{K}dr^\prime.
\end{equation}
For fully ionized hydrogen,
\begin{equation}
    n_e=n_{H^+}=\frac{\rho_{H^+}}{m_H}\sim \frac{\rho}{m_H}\,.
\end{equation}
Hence the optical depth of the fully ionized hydrogen gas is
\begin{equation}
    \tau(r) = \int_{r}^{+\infty}\frac{\rho(r^\prime)}{m_H}\sigma_{K}dr^\prime\,.
\end{equation}
The electron-photon cross section $\sigma_{K}$ in the non-relativistic regime can be approximated with the Thompson cross section, $\sigma_{\rm T}$.

\begin{figure}
    \includegraphics[width=.49\textwidth]{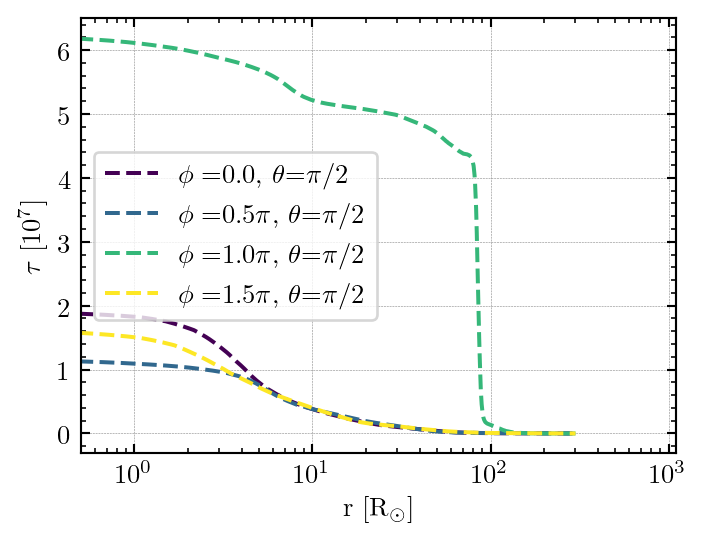}\\
    \includegraphics[width=.49\textwidth]{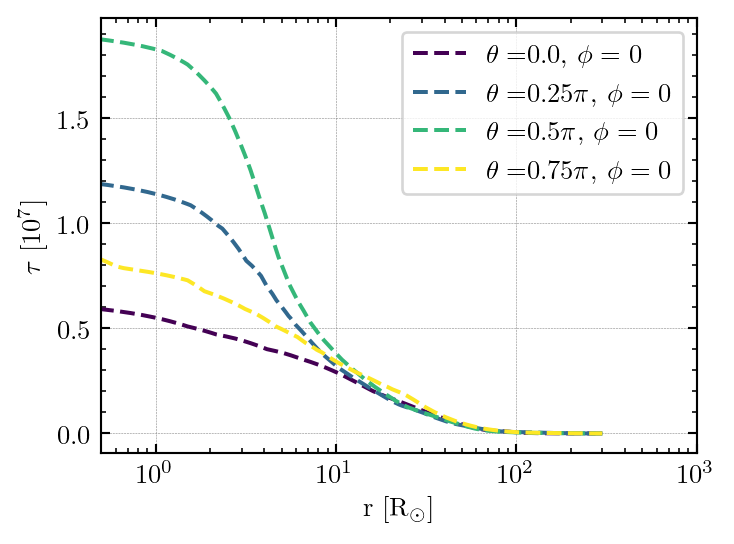}\\
    \includegraphics[width=.49\textwidth]{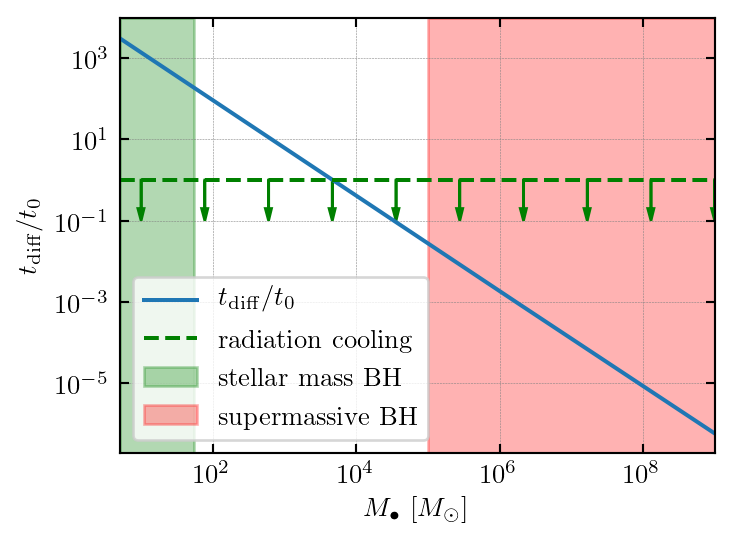}
    \caption{Optical depth of the accretion disk formed from a TDE by a 10~$M_\odot$ BH,  along the radial direction. \textit{Upper panel:} Optical depth for different values of the azimuthal angle $\phi$. \textit{Middle panel:} Optical depth for different values of the polar angle $\theta$. \textit{Bottom panel:} Ratio between the photon diffusion timescale and the dynamical timescale of the fallback material as a function of the BH mass. Radiative cooling of the disk  is efficient in the region below the dashed line, and hence only for BHs of masses $\gtrsim 10^4M_\odot$.}
   \label{fig:optical}
\end{figure}
The upper and middle panels of Figure~\ref{fig:optical} show the optical depth of the accretion disk in the radial direction for different azimuthal  and polar angles. As shown in this figure, the optical depth is on the order of $10^7$. By comparing with the value calculated from Equation~\ref{eq:tau} with geometry factor $\gamma=1$, we find that a more precise value for this factor is $\gamma\sim 30$. 

With such a large scattering optical depth, we assume the radiation is well thermalized and escapes with a roughly blackbody spectrum on a photon diffusion time scale, 
\begin{equation}
    t_{\rm diff} \sim \frac{\tau a_{\rm min}}{c}\,, 
\end{equation}
where $c$ is the speed of light.  The dynamical timescale of the most bound material is
\begin{equation}
    t_0 \sim \sqrt{\frac{a_{\rm min}^3}{G(M_\bullet+M_\star)}}.
\end{equation}
The bottom panel of Figure~\ref{fig:optical} shows the ratio between the diffusive  and the dynamical timescale as a function of the BH mass. It is evident that, unlike for the SMBH case, in the stellar BH case the timescale for photon diffusion is much longer than the dynamical timescale. Therefore, cooling is inefficient in the stellar mass BH regime, resulting in heating of the gas, and hence in a potentially dominant role of radiation pressure in the flow.

\subsection{Implementation of the Equation of State in   {\tt Phantom}}
\label{sec_implementation}

The SPH code {\tt Phantom} tracks 
shock heating when compiled
  without the ``isothermal" option; however, the central regions of the accretion disk -- where the gas is strongly shock-heated -- can reach very high temperatures,  making radiation pressure  non-negligible. Therefore, we implemented an equation of state that includes a radiation pressure term in {\tt Phantom}. In the following we discuss the details of this implementation.  

The equations of compressible hydrodynamics are solved in   {\tt Phantom} in the form of,
\begin{eqnarray}
\frac{d\mathbf{v}}{d t} &=&-\frac{\nabla P}{\rho} + \Pi_{\rm shock} + \mathbf{a}_{\rm ext} + \mathbf{a}_{\rm sink-gas} + \mathbf{a}_{\rm selfgrav}\\
\frac{d u}{dt} &=& -\frac{P}{\rho}(\nabla\cdot \mathbf{v}) + \Lambda_{\rm shock}- \frac{\Lambda_{\rm cool}}{\rho}\,,\label{eq:u-evol}
\end{eqnarray}
where $\mathbf{v}$ is the velocity field, $\mathbf{a}_{\rm ext}$, $\mathbf{a}_{\rm sink-gas}$ and $\mathbf{a}_{\rm selfgrav}$ refer to accelerations from an external force, sink particle and self-gravity, respectively. $\Pi_{\rm shock}$ and $\Lambda_{\rm shock}$ are dissipation terms required to  give the  correct entropy  increase at the shock front, and $\Lambda_{\rm cool}$ is the cooling term. 

\begin{figure}
    \includegraphics[width=.5\textwidth]{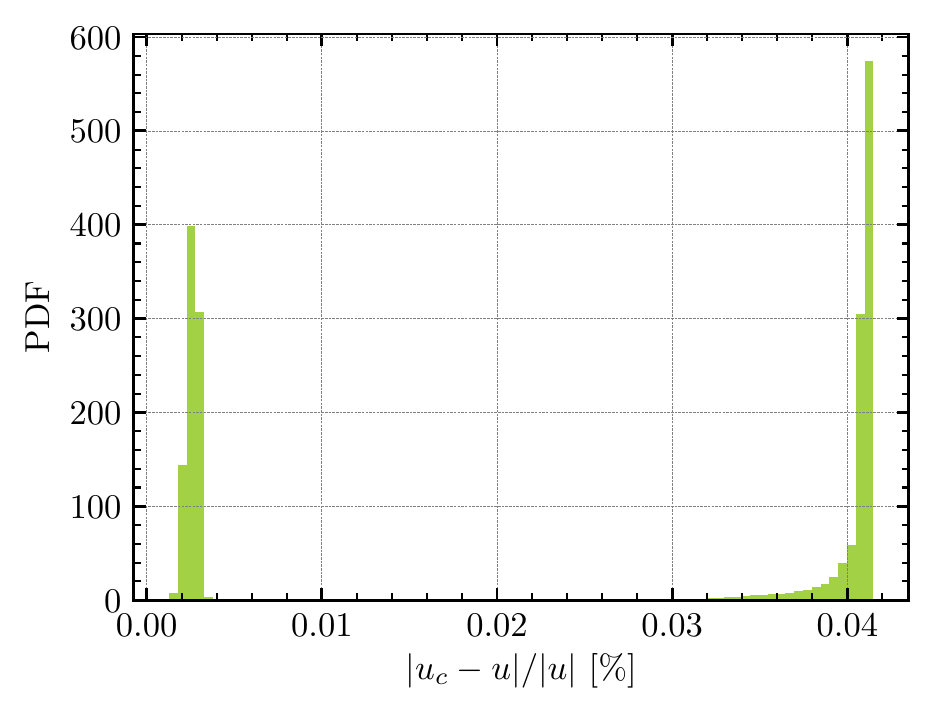}
    \caption{The distribution of the relative error of the specific thermal energy of all SPH particles in the implementation of the new EoS.  $u_c$ is the thermal energy calculated directly from the EoS in Equation~\ref{eq:u} with temperature obtained from the simulation, while $u$ is the corresponding thermal energy evolved via Equation~\ref{eq:u-evol}.}
    \label{fig:u-check}
\end{figure}

We implemented an equation of state including radiation pressure as,  
\begin{equation}\label{eq:u}
u = \frac{4\sigma}{\rho c}T^4 + \frac{k_{\rm B}(\gamma -1)}{\mu m_{\rm H}}T\,,
\end{equation}
where $u$ is the specific thermal energy, $\sigma$ is the Stefan-Boltzmann constant, $\rho$ the density of the SPH particle gas, $T$ the temperature, $k_{\rm B}$ the Boltzmann constant, $\gamma$ the adiabatic index, $\mu$ the mean molecular weight, and $m_{\rm H}$  the hydrogen mass.

At each time step, the current value of the specific thermal energy $u$ is used to calculate the temperature of each SPH particle via Equation~ (\ref{eq:u}) through Newton-Raphson iteration. The temperature in the last step is adopted as the initial guess of the temperature, while the density of the particle is obtained from the original computation in {\tt Phantom},
\begin{equation}
    \rho_a = \sum_b m_b W(|\mathbf{r}_a-\mathbf{r}_b|,h_a)\,,
\end{equation}
where $a$ and $b$ are particle labels, $m$ is the mass of the particle, $h$ is the smoothing length and $W$ the smoothing kernel. Hence the corresponding ratio of pressure over density, which is used to evolve the thermal energy, can be calculated from the updated temperature and density via
\begin{equation}\label{eq:ponrho}
\frac{P}{\rho} = \frac{4\sigma}{3\rho c}T^4 + \frac{k_{\rm B}}{\mu m_{\rm H}}T\,.
\end{equation}

Figure \ref{fig:u-check} shows the results of a test of the implementation of the modified equation of state (EoS). Specifically, 
in order to check that the variable $u$ representing the internal energy
is correctly evolved with the new equation of state (Equation~\ref{eq:u-evol}), we calculate
 the internal energy also in a different way, that is 
  from the stored temperature value $T$ in Equation~(\ref{eq:u}). We call this quantity $u_{\rm c}$, and check the consistency (i.e. the magnitude of the relative error) of $u$
  with respect to $u_c$.  Figure~\ref{fig:u-check} shows the distribution of the relative error of the specific thermal energy of all SPH particles. By setting the relative tolerance parameter of Newton Raphson iteration to be $10^{-6}$, the relative error on the specific thermal energy is under $10^{-4}$.

\subsection{Fallback rate of partial TDEs: powerlaws with the frozen-in approximation}
\begin{figure}
    \includegraphics[width=.49\textwidth]{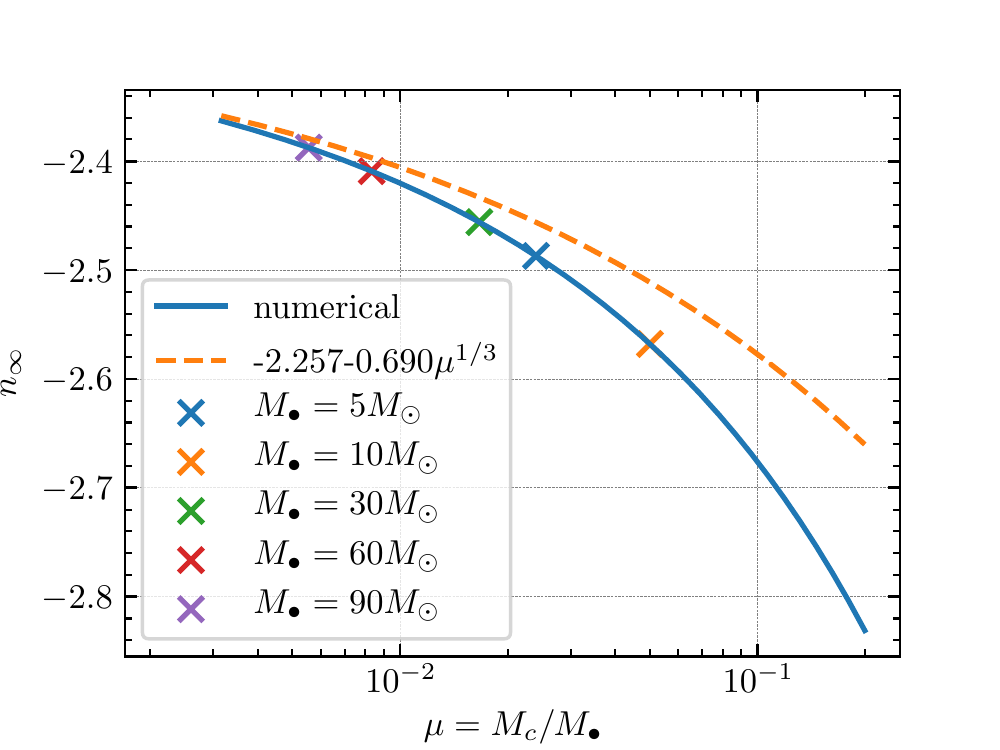}
    \caption{The powerlaw slope of the fallback rate for partial TDEs by stellar mass BHs. The value of the slope is 
    -2.409, -2.456, -2.568 and -2.676 for the BH masses 
    $M_\bullet=$5~$M_\odot$, 10~$M_\odot$, 30~$M_\odot$ and 60~$M_\odot$, respectively with a 0.5 $M_\odot$ remnant core. The value of the core mass is chosen based on what we find via SPH simulations (Sec.3.3). }
    \label{fig:n_infty}
\end{figure}

To estimate the fallback rate, the so-called ``frozen-in" approximation  \citep{Rees1988,Phinney1989,Ulmer1999} is often adopted; according to this,
\begin{equation}\label{eq:53}
    \frac{dM}{dt}=\frac{dM}{dE}\frac{dE}{dt}=\frac{1}{3}(2\pi G M_\bullet)^{2/3}\frac{dM}{dE}t^{-5/3}\,,
\end{equation}
where $M_{\bullet}$ is the BH mass. By calculating the $dM/dE$ distribution  of the gas particles from the snapshots of the simulation, one can then estimate the fallback rate as a function of time.  In this approximation, the accretion rate is calculated from the time that the debris take to return to the pericenter. It requires the following conditions to be satisfied: (i) the viscous timescale of the disk be much shorter than the fallback time; (ii) the kinetic energy of the falling back material dissipate efficiently; and (iii) the material can be rapidly circularized.  For full tidal disruption by SMBHs, the specific energies of the debris can approximately be taken to be "frozen-in" at the tidal radius; thus, the fallback rate will asymptote to $t^{-5/3}$ at late times. However, for partial disruption by stellar mass BHs, where the remnant core provides additional significant time-dependent potential to the BH potential, the fallback rate will be steeper than $t^{-5/3}$. \citet{Coughlin2019} developed a model to analytically calculate the asymptotic scaling of the fallback rate at late times from a partial TDE. By solving the equation of motion of the particles in the stream together with the continuity equation, they found that the fallback rate can be expressed as,
\begin{equation}
    \dot{m}\propto t^{-1-\frac{2}{3}\omega_+} =t^{n_\infty}\,,
    \label{eq:mdot}
\end{equation}
where the variable $\omega_+$
is derived from the equations,
\begin{eqnarray}
    \omega_+ &=& \frac{1}{4}\bigg( -1+\sqrt{9+\frac{16}{\xi_\infty^3}+\frac{16\mu}{(1-\xi_\infty)^3}}\bigg)\\
    \xi_\infty&=&\frac{1}{\xi_\infty^2}-\frac{\mu}{(1-\xi_\infty)^2},\label{eq:xi}
\end{eqnarray}
where $\mu\equiv M_c/M_\bullet$ is the mass ratio between the remnant core and the BH,  and $\xi_\infty = r(t\rightarrow\infty)/R(t\rightarrow\infty)$, where $r(t\rightarrow\infty)$ is the distance between the BH and the test particle and $R(t\rightarrow\infty)$ is the distance between the remnant core and the test particle. The symbol '$\infty$' is to remind us that this is the asymptotic time.

For full tidal disruption, where $\mu=0$, $\xi_\infty = 1$, $\omega_+ = 1$, Equation~\ref{eq:mdot} yields the classic $\dot{m}\propto t^{-5/3}$. For partial disruption by a SMBH, where $\mu \ll 1$, $\xi_\infty$ can be approximated by
\begin{equation}
    \xi_\infty \sim 1-\left(\frac{\mu}{3}\right)^{1/3}+\frac{3}{8}\left(\frac{\mu}{3}\right)^{2/3}+\mathcal{O}(\mu)
\end{equation}
which yields
{
\begin{equation}
    \dot{m}\propto t^{-2.257-0.690\mu^{1/3}} \,.
\end{equation}
}
Since for tidal disruption by a SMBH the ratio $\mu$ is always $\ll1$ and insensitive to the mass of the core,  they conclude that for  partial TDEs by SMBHs the fallback rate will asymptote to $t^{-2.257}\sim t^{-9/4}$ and is effectively independent of the mass of the remnant core that survives the encounter. { \citet{Coughlin2019} solved Equation~(\ref{eq:xi}) numerically within the range $\mu\in$[$10^{-9}$,$10^{-5}$] and found that the solution agrees well with the $t^{-9/4}$ decay.} However, for TDEs by stellar mass BHs, the approximation $\mu \ll1$ no longer holds true; {thus, numerical solution of Equation~(\ref{eq:xi}) is needed to obtain a more accurate value of $\xi_\infty$ { (Note that \citealt{Coughlin_2020} provided an example simulation where $\mu \approx$ 0.01 )}.}

Figure~\ref{fig:n_infty} shows the corresponding  powerlaw index of the fallback rate, 
$n_\infty$, computed  from the numerical solution of $\xi_\infty$ as a function of $\mu$. The calculation assumes a mass of the remnant core $M_{\rm c}=0.5M_\odot$, chosen to match what is found from the SPH simulations presented later (Section~\ref{sec_fallback}).

We see that in the stellar mass regime, $n_\infty$ changes significantly with the mass of the BH. For  masses from 
about $5~M_\odot$ up to $90~M_\odot$, $n_\infty$ ranges from $\sim$ -2.4 to $\sim$ -2.7. Therefore, for partial TDEs by stellar mass BHs, due to the time-dependent impact of the remnant core onto the falling back material, the power law slope of the fallback rate shows a significant dependence on the BH mass.  

\subsection{Fallback rate calculation}
\begin{figure*}
  
    \includegraphics[width=.3\textwidth]{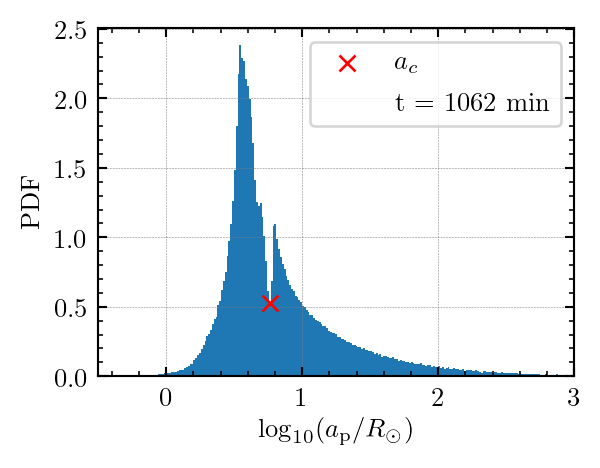}
    \includegraphics[width=.3\textwidth]{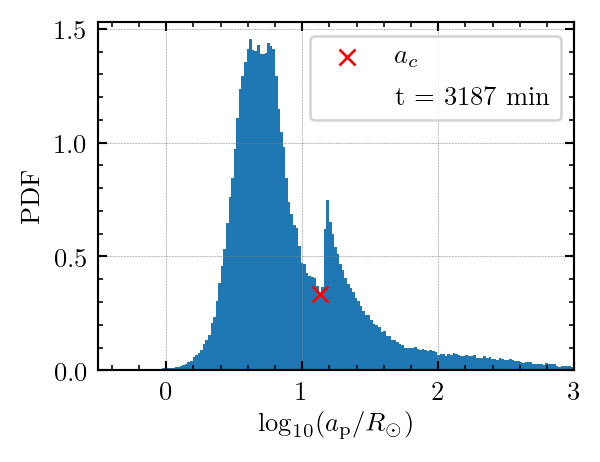}
    \includegraphics[width=.3\textwidth]{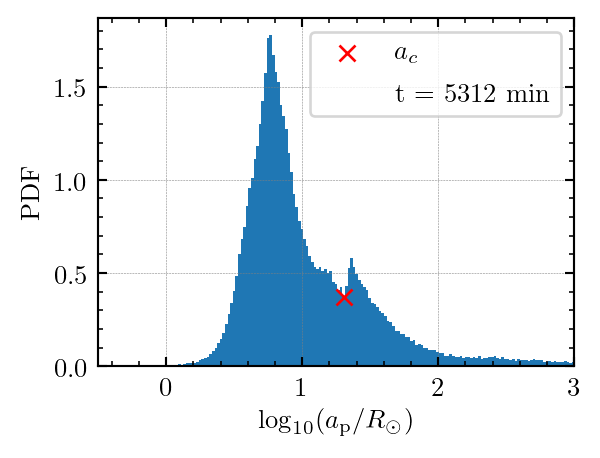}\\
    \includegraphics[width=.3\textwidth]{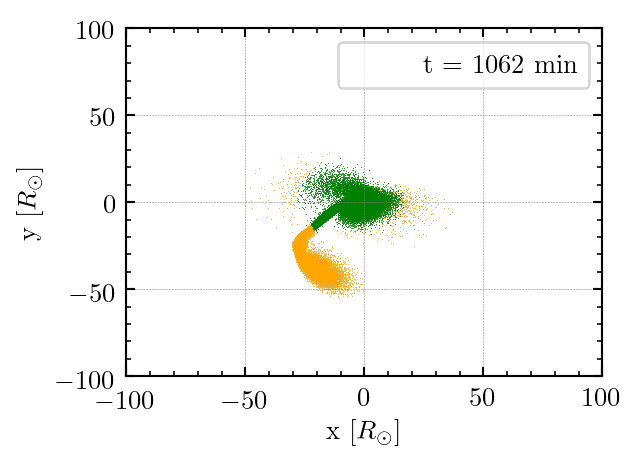}
    \includegraphics[width=.3\textwidth]{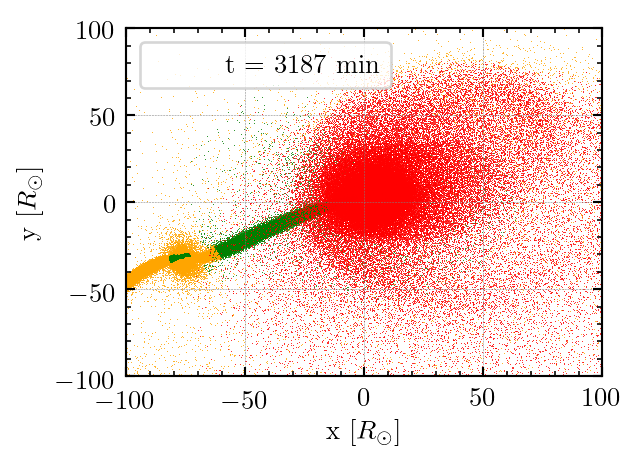}
    \includegraphics[width=.3\textwidth]{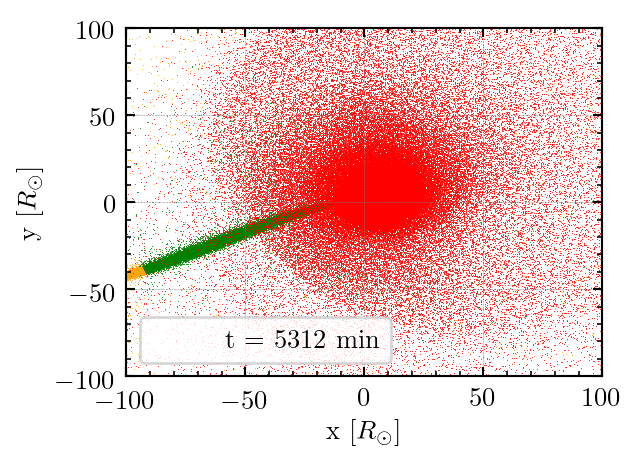}
    \caption{The upper three panels show the distribution of the semi-major axis of the SPH particles around the BH at three different times. A clear separation can be seen between the disk particles (left peak with small semi-major axis) and the stream particles (right peak with large semi-major axis). The bottom three panels show the corresponding snapshots with the SPH particles color coded based on this direct classification method. The red color indicates the disk particles which have circularized, the green color indicates the bound particles falling back, while the yellow shows the unbound particles. The fallback rate can then be estimated by $\dot{M}_{\rm disk}$. Note that some of the particles bound to the BH (green) may later become unbound (orange) due to the presence of the remnant core or join the disk (red) and subsequently accrete. }
    \label{fig:a-dist}
\end{figure*}

As discussed in the previous subsection, the fallback process of the debris for TDEs by stellar mass BHs is more complicated than for full TDEs by SMBHs (which yield the traditional $-5/3$ powerlaw slope), due to the existence of a remnant core of non-negligible mass compared to that of the BH. 
To calculate the fallback rate from the simulation, traditional methods usually extract the quantity ${dM}/{dE}$ from the snapshots, and then compute the fallback rate according to the frozen-in approximation (Equation \ref{eq:53}). In order to estimate ${dM}/{dE}$ from the snapshots, the 
first step is a calculation of the total specific energy of the SPH particles with
 respect to the black hole, of mass $M_\bullet$,
\begin{equation}\label{eq:etot}
    E = \frac{v^2}{2}-\frac{GM_\bullet}{r}\,,
\end{equation}
and then a sorting on all SPH particles to determine the distribution ${dM}/{dE}$.  
{ While this method is still used in the literature},
it becomes invalid in the case of partial TDEs, and especially so for TDEs by a stellar mass BH. Because of the presence of the remnant core, the material in the stream no longer obeys the Keplerian law when falling back to the pericenter. Hence the underlying assumption leading to Equation~(\ref{eq:53}) is no longer correct. Therefore, a more precise method that can handle this case is required and 
{to this effect, several alternative approaches have been employed in the last few years which measure the accretion rate directly \citep{Coughlin2017,Coughlin_2020}}.

In this paper we propose another new direct method to estimate the fallback rate: we classify the SPH particles directly from the snapshots into disk particle, accreted particles, stream particles and remnant particles.

During  accretion, the bound particles continuously fall back on the BH. However, the particles in the stream are nearly parabolic with extremely large semi-major axis, while the particles in the accretion disk have circularized following collision with the stream. The semi-major axis and the eccentricity of the individual particles are calculated from 
\begin{eqnarray}\label{eq:ae}
    a &=& -\frac{GM_\bullet r}{rv^2-2GM_\bullet}\\ 
    e &=& \frac{rv^2-GM_\bullet}{GM_\bullet}\,,
\end{eqnarray}
where $r$ and $v$ are the absolute values of the relative position and velocity between the SPH particles and the BH.

Due to self-collision,  circularization of the accreted material will create a natural separation in the distribution of the semi-major axis/eccentricity of the bound particles, where all the particles with small semi-major axis/small eccentricity belong to the disk, while all the particles with large semi-major axis/large eccentricity are part of the stream/remnant. If we add the mass of the disk particles by snapshots, we thus obtain the fallback mass as a function of time. Then the fallback rate can be readily obtained from the derivative without any approximation. Figure \ref{fig:a-dist} shows the distribution of the semi-major axis in the particles in several snapshots,  together with the corresponding results of the classification method (note that the accreted particles have not been indicated since they are added to the BH mass and no longer evolved). The upper three panels show the distribution of the SPH particles which remain bound to the BH, at different times after the first encounter. We can see a clear bimodality in the distribution of the semi-major axis of the falling-back debris, with a sharp separation between the two peaks. This is because  circularization rapidly decreases the semi-major axis once the material falls back to the BH. Therefore, we can easily distinguish the SPH particles that have been accreted. Based on these findings, we compute the  fallback rate as follows:

\begin{enumerate}
   \item Identify all the SPH particles that have been removed from the simulations due to accretion by the BH and mark them as accreted particles.
   \item For each of the non-accreted particles, calculate the specific energy via Equation~(\ref{eq:etot}); for particles with $E>0$, mark them as unbound particles, for particles with $E<=0$, mark them as bound particles.
   
   \item For all bound state particles, calculate the semi-major axis of each particle with respect to the BH via Equation~(\ref{eq:ae}).
   
   \item Find the critical value of the  semi-major axis, $a_c$ (by means of a peak/valley finding algorithm)  that separates stream particles from disk particles. For particles with $a<a_c$, mark them as disk particle, while for particles with $a>a_c$, mark them as stream particles.
   
   \item Add up the mass of all disk particles and accreted particles to obtain the total mass of fallback material.
   
   \item Repeat the process by snapshot to get the fallback mass as a function of time, $M_{\rm fallback}(t)$.
   
   \item Derive the fallback accretion rate $\dot{M}_{\rm fallback}(t)$
   by taking the numerical derivative 
   of $M_{\rm fallback}(t)$.
\end{enumerate}

The bottom three panels of Figure \ref{fig:a-dist} show the resulting particle classification by means of our method. In red are disk particle, in green are stream particles, while orange depicts unbound particles. The figure shows that the particles are successfully classified into these three kinds. Therefore, with this method, a more accurate fallback rate can be extracted from the simulations, without relying on the frozen-in approximation. 

\subsection{Simulation setup}\label{sec:setups}

In the following we describe our simulation setup in order to study the fallback rate of partial TDEs by stellar mass BHs with   {\tt Phantom}.
This code has been highly effective for simulating complex fluid geometries \citep{Nixon2012, Nixon2013,Martin2014a,Martin2014b,Nealon2015,Dogan2015}, and has also been used to study TDEs, including the disruption process itself \citep{Coughlin2015}, the evolution of the disrupted debris \citep{Coughlin2016}, and the formation of the accretion disk \citep{Bonnerot2016}.
For the simulations in this paper, we adopt an artificial viscosity,
$\alpha^{\rm AV}$, varying between the range  $\alpha_{\rm min}^{\rm AV}=0.1$ and $\alpha_{\rm max}^{\rm AV}=1$.  

The disrupted star is initialized as an {$n=3/2$ polytrope of one solar mass and one solar radius, while stellar mass BHs of 5, 10, 30, 60 and 90 solar masses are considered. The sizes (accretion radii) of the BHs are set to be 2000, 1000, 333, 167 and 111 times larger than their event horizons for practical computational time. Any SPH particle entering within the sink radius is accreted onto the BH; its mass is added to that  of the BH and it is removed from the simulation.  All stars are assumed to be incident with penetration factor equal to 1 from an initial position of 50 tidal disruption radii. The gravitational field of the BH is purely Newtonian and treated as a sink particle in {\tt Phantom}; thus general relativistic  precession is not included in our simulations.

Each simulation is performed with $5\times 10^5$ SPH particles. We checked the convergence of the results by performing one simulation also with $10^6$ SPH particles. Note that self-gravity of the gas is included in the simulations. Both analytical and numerical studies have demonstrated that self-gravity plays an important role for modifying the structure of the disrupted debris \citep{Kochanek1994,Coughlin2016,Coughlin2015,Coughlin2016a}, especially when the pericenter distance of the disrupted star is larger ($\beta \sim 1$) and the adiabatic index of the gas is stiffer ($\gamma>5/3$) \citep{Coughlin2016}. The modification of the structure from  self-gravity becomes more significant in the case of tidal disruption by a stellar mass BH.  

Three different setups are explored in our simulations to study the impact of shock heating and radiation from  accreting material. The first one is the default setting in {\tt Phantom} for TDE simulations, that assumes an adiabatic equation of state with index equal to the polytropic value of $\gamma = 5/3$  (the ``isothermal" compiling option). This setup assumes that the gas cools very efficiently after being shocked, and the temperature remains low enough that the contribution of radiation to the gas dynamics is small. Both shock heating and radiation would contribute to inflate the gas and extend the accretion regions. The second setting  we consider is again the adiabatic equation of state with index equal to the polytropic value of $\gamma = 5/3$ but without the ``isothermal" condition. This setup includes heating from the PdV work and shock heating and does not assume that the gas cools very efficiently after being shocked. Therefore, the temperature may not remain low enough and the contribution from the radiation needs to be taken into account. The last setup, where the radiation is fully included, is the one with our implementation of the modified equation of state as discussed in Section~\ref{sec_implementation}. With this equation of state the temperature is updated in each step and the influence of radiation pressure (but no other radiative transfer effects) on the gas dynamics is accounted for. While the third setup is the one most physically accurate for the problem we are treating, the other two cases are useful for comparison and to better appreciate the importance of including both shock heating and radiation pressure in the simulations.

\section{Results}
\label{sec:results}
\subsection{Morphological evolution}

\begin{figure*}
      \includegraphics[width=.33\textwidth]{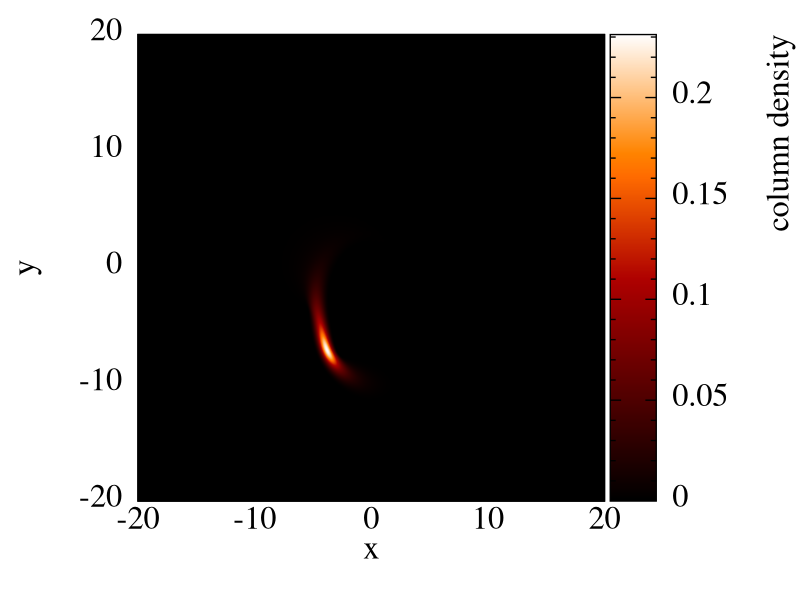}
      \includegraphics[width=.33\textwidth]{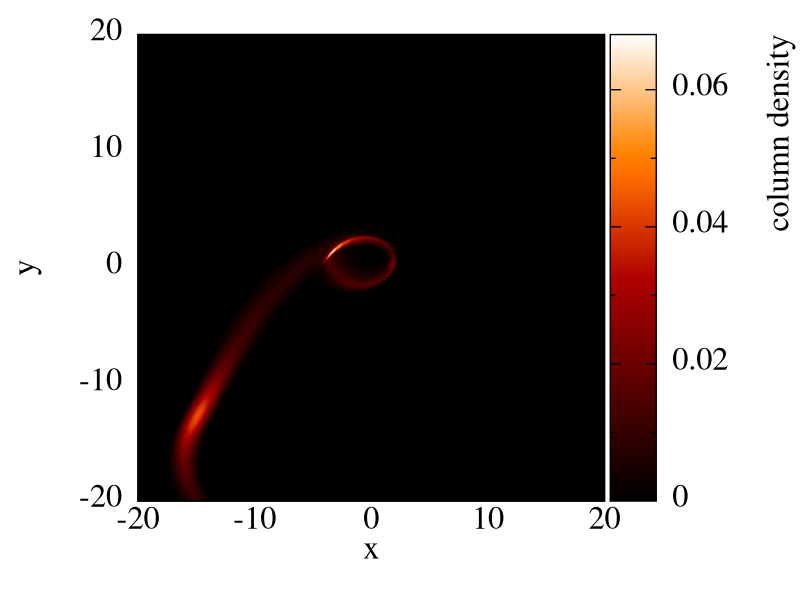}
      \includegraphics[width=.33\textwidth]{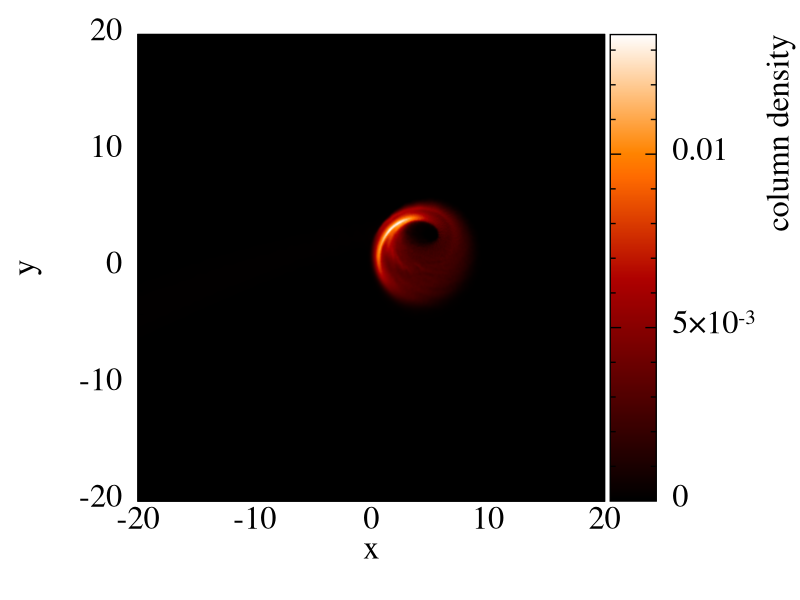}\\
    \caption{Snapshots of the simulation of the tidal disruption of a 1~$M_\odot$ star by a 10~$M_\odot$ BH; an adiabatic equation of state with $\gamma=5/3$ is adopted. Each panel shows the column density projection on the x-y plane at $t = 3.2\times 10^4$~s, $t = 4.8\times 10^4$~s and $t = 2.56\times 10^5$~s, respectively. }
    \label{fig:iso-snaps}
\end{figure*}

Figure \ref{fig:iso-snaps} shows the projection in the x-y plane (defined as the orbital plane of the incoming star)  of the evolution of the debris. In this simulation,   $M_\bullet= 10~M_\odot$, $M_* = 1~M_\odot$, penetration number $\beta = 1$, and the adiabatic EoS is used under the isothermal assumption, where cooling of the gas is assumed to be very efficient after the shock. Since this is the default setup of   {\tt Phantom} for TDEs, we use this as the standard simulation for comparison. The three panels of Figure~\ref{fig:iso-snaps} show snapshots of the TDE at the times $t = 3.2\times 10^4$~s, $t = 4.8\times 10^4$~s and $t = 2.56\times 10^5$~s after the first encounter, respectively. In these panels, the color shows the projected column density of the gas, with lighter tones corresponding to denser areas. At the beginning of the simulation, the center of mass of the BH-star system is placed at the origin. However, because of the relatively  small mass of the stellar BH, the BH obtains a recoil velocity from the encounter with the star. As we can see from the bottom two panels of Figure~\ref{fig:iso-snaps}, the BH, accompanied by the accretion disk, gradually moves away from the origin. To zeroth order, if we ignore the self-gravity of the stream, the dynamics of the falling back material is determined by its relative position with respect to the BH and the remnant core. Therefore, the recoil velocity of the BH might in principle affect the fallback rate of the TDE if a remnant core is present (the motion of the falling back material is non-Keplerian). To explore the significance of this dependence, we performed a comparison simulation with $M_\bullet=10~M_\odot$ that forces the BH to remain  at its initial position, finding that there is no significant difference due to the recoil velocity. Larger mass BHs will obtain smaller recoil velocity; therefore, we conclude that the recoil velocity obtained from the encounter between a 1~$M_\odot$ star and a stellar mass BH (with $M_\bullet \gtrsim 10M_\odot$) does not significantly affect the accretion process. However, close encounters between massive stars and stellar mass compact objects are possible in clusters. In close encounters between objects of comparable mass, the compact object will get a much higher recoil velocity that might significantly affect the accretion process. We leave the exploration of the impact from the recoil velocity to future work. 

As discussed earlier, the default setup for TDE simulations  in   {\tt Phantom} assumes that shock heating of the gas is radiated efficiently after the gas is shocked via stream-stream collisions. Numerically, this means that the polytropic constant remains unaffected, and hence shock heating does not contribute to the thermal balance and cannot ``inflate" the accretion disk \citep{Coughlin2017, Hayasaki2016}, which thus remains confined within a few tidal radii in the radial direction and stays geometrically thin. Moreover, due to the high accretion rate (which can be two orders of magnitude higher than the Eddington value), the radiation pressure cannot be ignored. \citet{Coughlin2014} found that radiation pressure can be very important in expelling gas and inflating the accretion disc. Therefore, shock heating and radiation pressure need to be included in the simulations. We proceeded in steps of generalization.
First, we performed the same TDE simulation with the polytropic EoS but without enforcing isothermal conditions in   {\tt Phantom}. This run allows us to correctly capture shock heating. Then, we performed another simulation with the modified equation of state, where radiation pressure is added to the thermal dynamics as detailed in Section~\ref{sec_implementation}.

\begin{figure*}
      \includegraphics[width=.32\textwidth]{iso/splash_0160.png}
      \includegraphics[width=.32\textwidth]{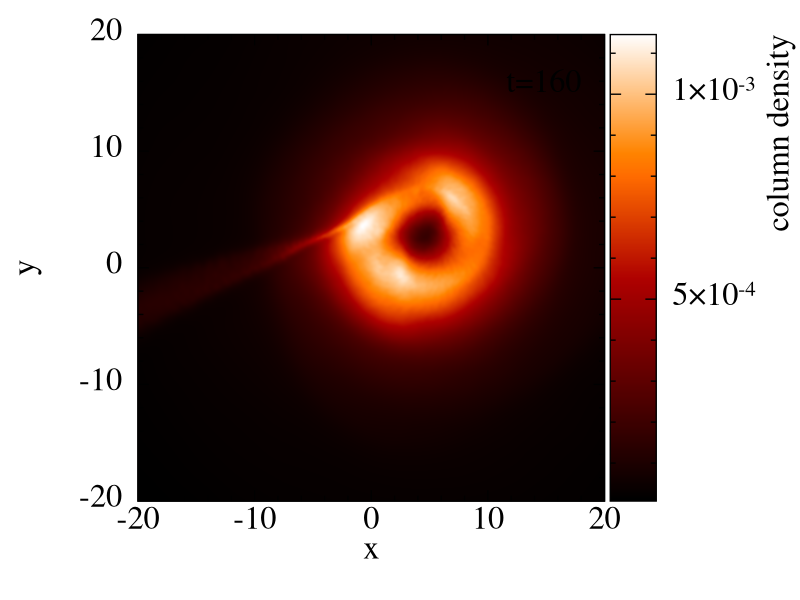}
      \includegraphics[width=.32\textwidth]{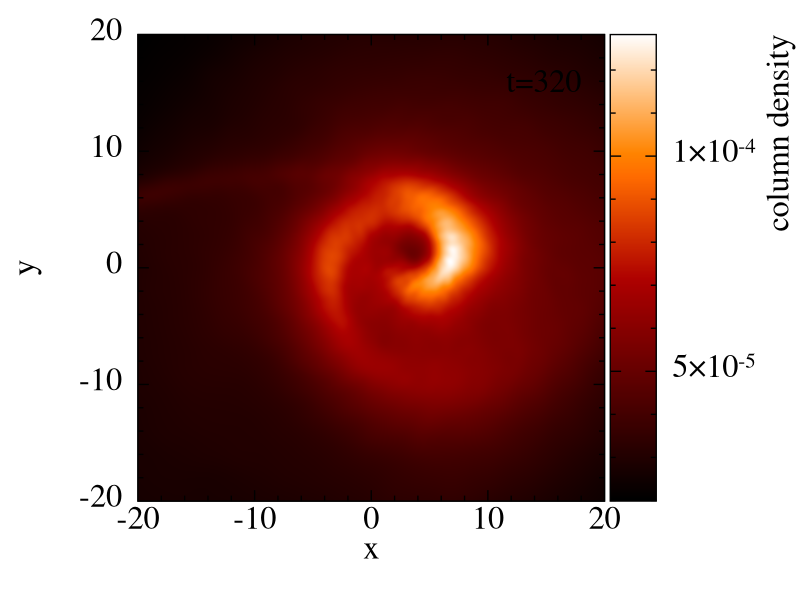}\\
      \includegraphics[width=.32\textwidth]{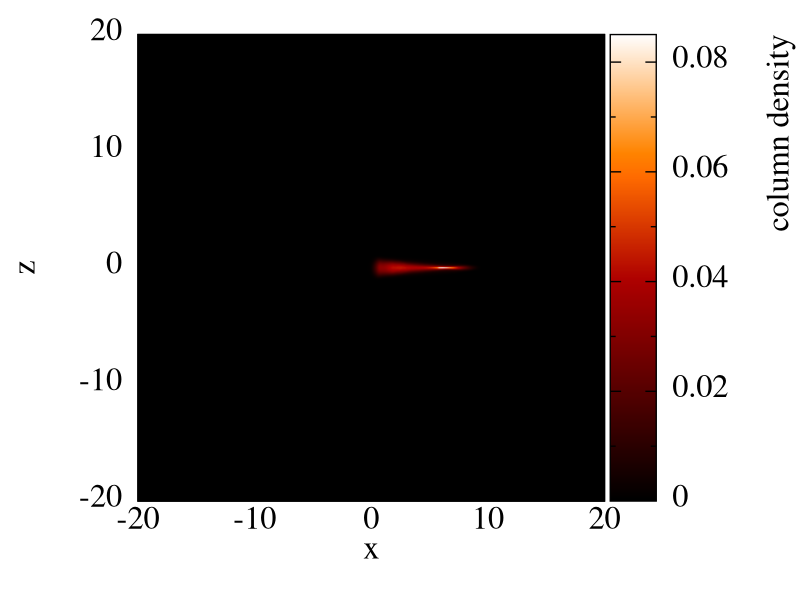}
      \includegraphics[width=.32\textwidth]{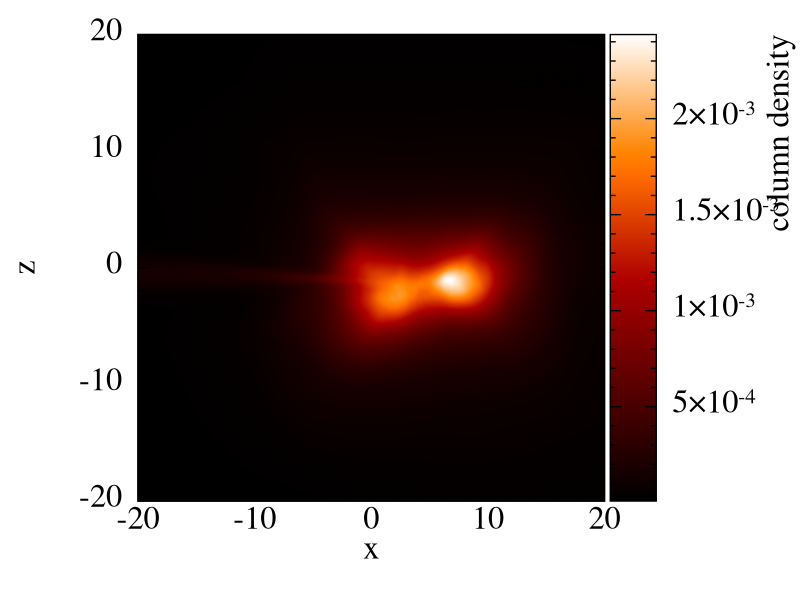}
      \includegraphics[width=.32\textwidth]{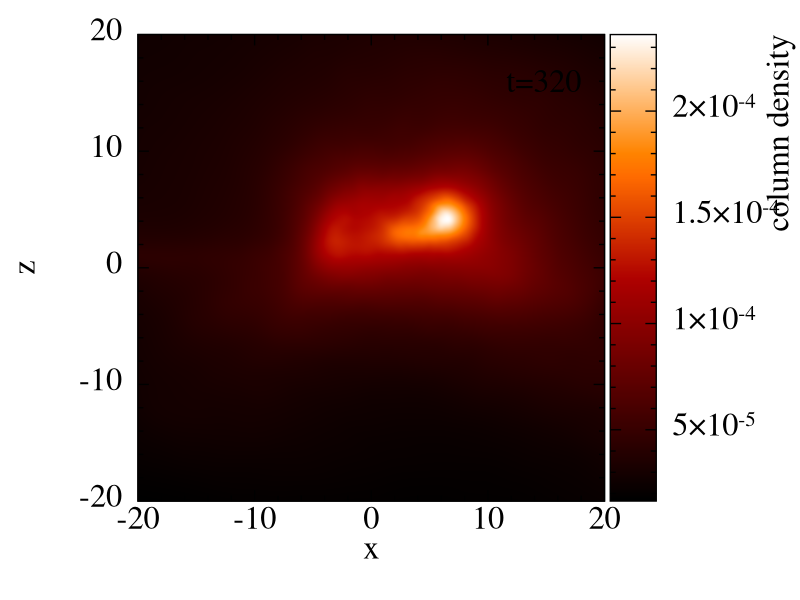}\\
    \caption{Projection of the accretion disk structure in the x-y plane (upper panels) and in the x-z plane (bottom panels) from three different simulations but all at the same time $t=2.56\times 10^5$~s for a TDE of a 1~$M_\odot$ star disrupted by a 10~$M_\odot$ BH. \textit{Left}: Default setup of   {\tt Phantom}, where neither shock heating nor radiation pressure are included. \textit{Middle}:  Shock heating is correctly captured but radiation pressure is not included. \textit{Bottom}: Both shock heating and radiation pressure are included.}
    \label{fig:disk-comp}
\end{figure*}

Figure \ref{fig:disk-comp} shows the structure of the accretion disk from the three different setups. The snapshots are taken at $t = 2.56\times 10^5$~s after the encounter. The upper three panels show the projection in the x-y plane, while the bottom three panels show the corresponding projection in the x-z plane. The first column is the default setup of the TDE in   {\tt Phantom} where we see a confined, thin disk. In the second column shock heating is captured, and the accretion disk is ``puffed up" by the 
shock heating from stream-stream collisions. We observe that a more circular and thicker disc is formed. The third column shows the accretion disk from the simulation where both shock heating and radiation pressure are included. Here the accretion disk is inflated to a larger size due to the effect of the radiation pressure. We also found that the gas in the disk is continuously expelled outward to space.

\begin{figure*}
      \includegraphics[width=.49\textwidth]{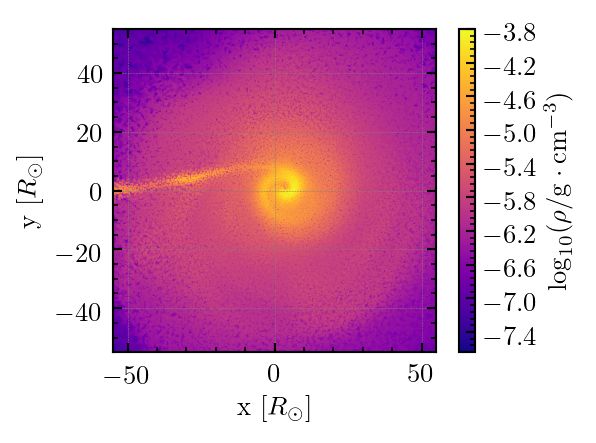}
      \includegraphics[width=.49\textwidth]{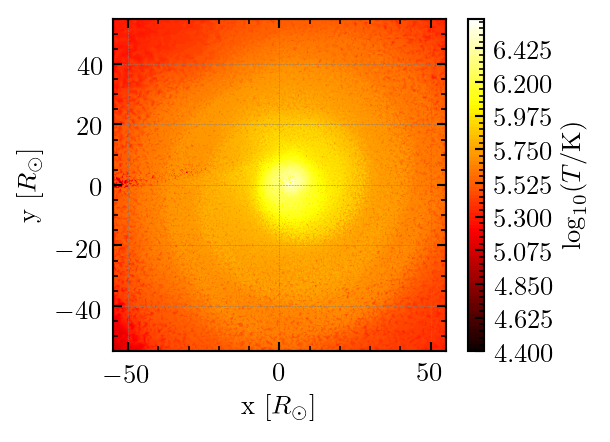}\\
      \includegraphics[width=.49\textwidth]{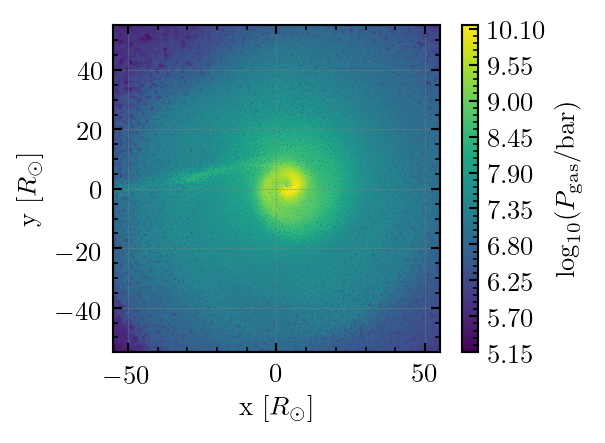}
      \includegraphics[width=.49\textwidth]{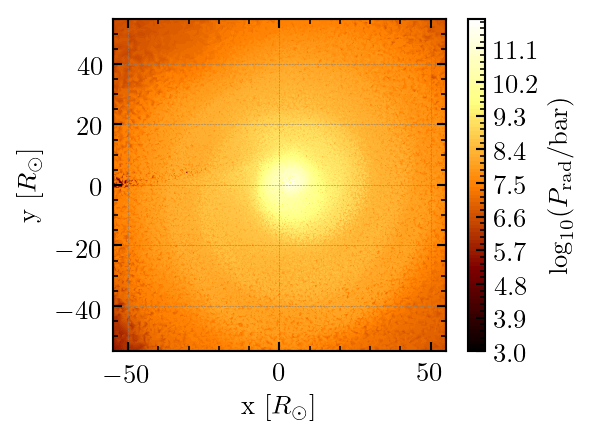}
    \caption{Column density (\textit{upper left}), temperature (\textit{upper right}), gas pressure (\textit{bottom left}) and radiation pressure (\textit{bottom right}) of the accretion disk at time $t=8500$~min from a simulation with shock heating and radiation pressure included. }
    \label{fig:prop}
\end{figure*}
Figure~ \ref{fig:prop} shows the column density, temperature, gas pressure and radiation pressure of the accretion disk corresponding to the third column of Figure~\ref{fig:disk-comp}. We see that  radiation pressure (bottom right panel) dominates over gas pressure in the central regions of the disk. Here the radiation pressure exceeds the gas pressure by approximately a factor of ten. This finding confirms the fact that, to correctly resolve the accretion disk structure of the TDE by stellar mass BHs, the inclusion of radiation pressure in the equation of state is necessary. 

\subsection{Stream instability}
In the stream, to a first approximation, self-gravity dominates the tidal field of the BH when
\begin{equation}
    \rho \gtrsim \frac{M_\bullet}{r^3}\,,
\end{equation}
where $\rho$ is the local density of the gas and $r$ is the distance to the BH. By considering the dynamics of the stream post-disruption, \citet{Coughlin2016a} found that compression and the onset of significant stream self-gravity occurs more readily for lower-mass black holes. 
 If any overdense region with size $\delta R$ is perturbed and the Jeans condition is satisfied, 
\begin{equation}
    \delta R \gtrsim c_s\sqrt{\frac{\pi}{\rho G}},
\end{equation}
where $c_s$ is the sound speed, the material will collapse into a clump. \citet{Coughlin2015} discussed this effect in the SMBH regime. They argued that the stream is gravitationally unstable, and any small perturbation to the distribution of the debris will cause the stream to  fragment. They argued that those collapsed fragments significantly affect the  fallback rate, creating varied, abruptly variable light curves. Here, due to the relatively weaker gravity of the stellar mass BH, we argue that the stream is much more gravitationally unstable than the stream in TDE by SMBHs.
 We find in the simulations that if  shock heating and radiation pressure are correctly included, the stream almost entirely collapses into a series of clumps. This significantly affects the fallback rate, and subsequently induces a series of abrupt changes in the light curves.

\begin{figure}
      \includegraphics[width=.49\textwidth]{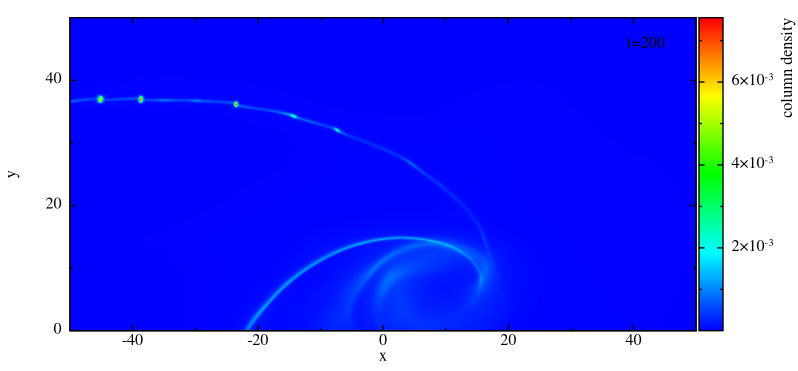}\\
      \includegraphics[width=.49\textwidth]{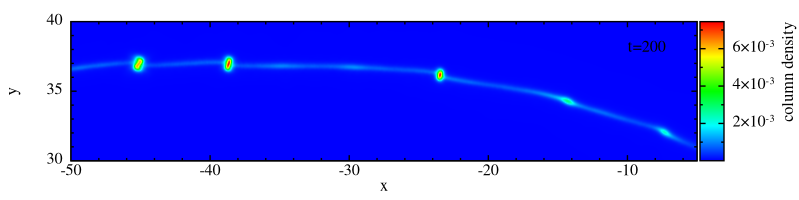}\\
    \caption{Stream instability in a partial TDE by a BH of  5~$M_\odot$. The material in the stream collapses into a series of clumps before falling back to the BH. The bottom panel shows a zoom-in view of the clumped stream that is seen in the upper panel.}
    \label{fig:stream}
\end{figure}

Figure~\ref{fig:stream} shows a snapshot at time {$t=3.2\times10^5$~s} from one of our simulations, that is the tidal disruption of a 1~$M_\odot$ star by a 5~$M_\odot$ BH, with shock heating and radiation pressure included. From the upper panel and a zoomed region in the bottom panel, we can clearly see the collapsed clumps by eye.  Unlike the clumps in TDEs by SMBHs shown in \citet{Coughlin2015}, the central density of the clumps is much denser than the nearby areas. Therefore, the abrupt changes in luminosity caused by those clumps are expected to be much more prominent.

\begin{figure}
      \includegraphics[width=.49\textwidth]{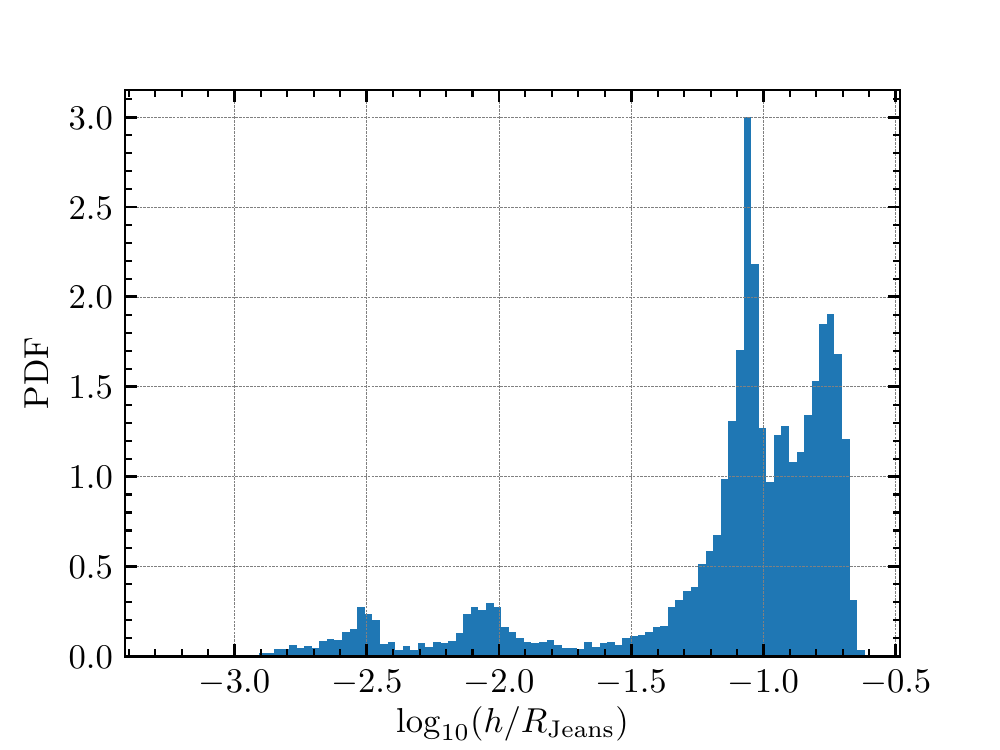}\\
    \caption{The distribution of the smoothing length of the SPH particles in unit of Jeans radius.}
    \label{fig:Jeans}
\end{figure}

To make sure that we correctly resolve the collapse of the stream, we need to test that the smoothing length of the SPH particles is smaller than the Jeans radius $R_{\rm Jeans}=c_{s}\sqrt{\frac{\pi}{G\rho}}$. Figure~\ref{fig:Jeans} shows the distribution of the smoothing length of the SPH particles in the bottom panel of Figure~\ref{fig:stream}, in units of $R_{\rm Jeans}$. We see that all particles have smoothing length smaller than the Jeans radius. Therefore, the collapse is correctly resolved by our simulations.

\subsection{Fallback rate}
\label{sec_fallback}
As discussed earlier, due to the small cross section for full TDEs by stellar mass BHs, most of the disruptions in dense clusters in which close encounters are frequent will result in partial disruptions. The presence of the remnant stellar core imposes a significant time-dependent potential on the falling back stream. In the SMBH regime, where the mass ratio between the remnant core and the BH is very small, the late time fallback rate asymptotes to $t^{-9/4}$ and is independent of the mass of the BH. However, in the stellar mass regime, where the mass ratio $\mu$ is several orders of magnitude higher than in the SMBH regime, the fallback rate shows a noticeable dependence on the mass of the BH.

\begin{figure}
      \includegraphics[width=.49\textwidth]{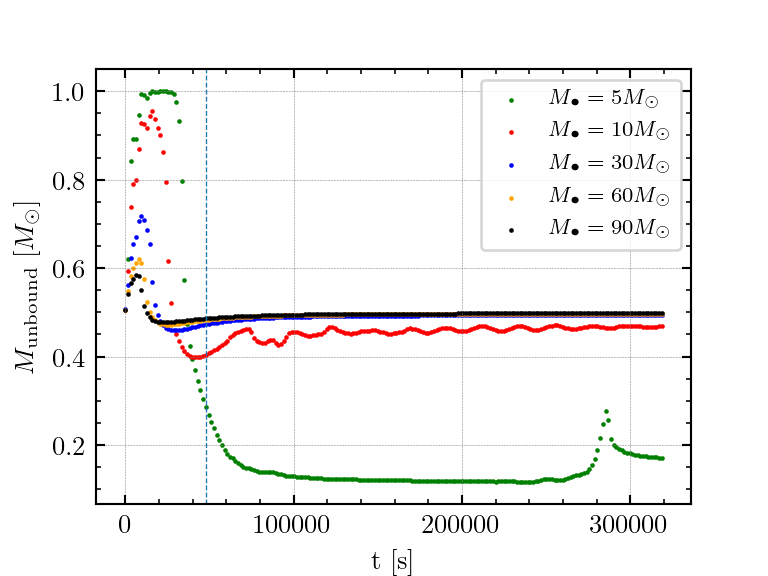}\\
      \includegraphics[width=.46\textwidth]{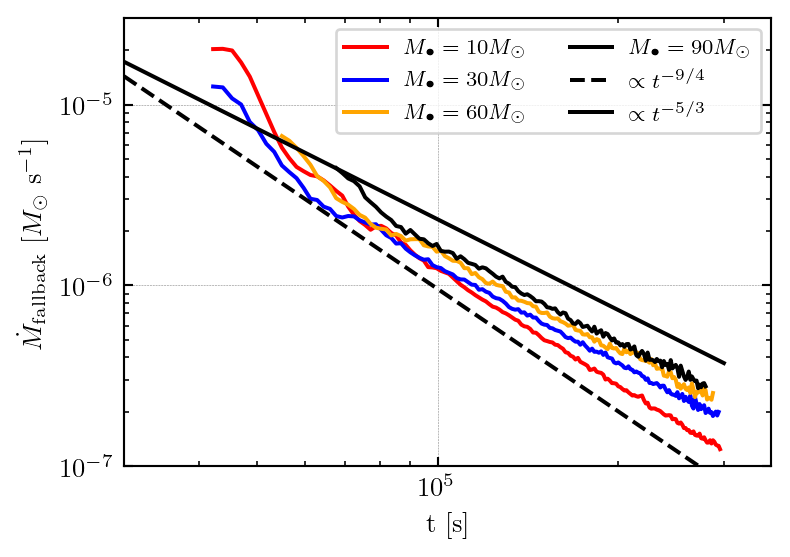}\\
      \includegraphics[width=.46\textwidth]{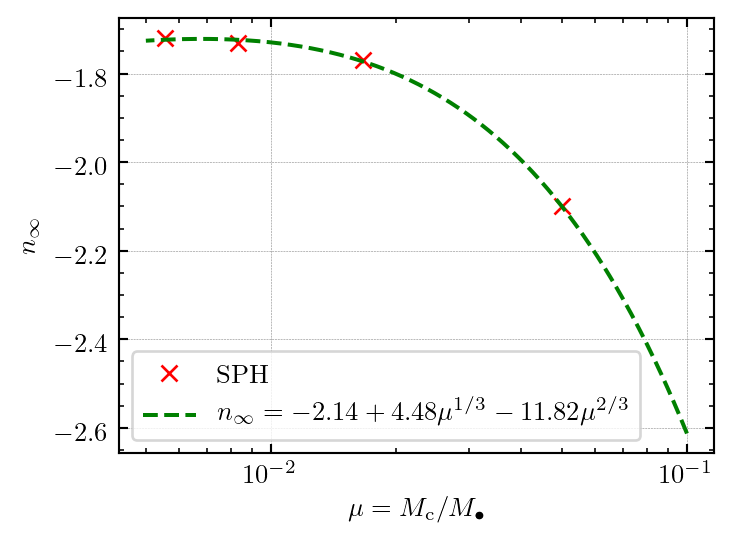}
    \caption{The mass of the unbound material (\textit {upper panel}) and the fallback rate (\textit{middle panel}) as a function of time, in TDE simulations with BH mass equal to 5, 10, 30, 60 and 90 $M_\odot$. The late-time
    powerlaw slopes of the fallback rate for 10, 30, 60 and 90 $M_\odot$ are -2.10, -1.77, -1.73 and -1.72, respectively.  \textit{Bottom panel:} fit results for the late time slope of the fallback rate  as a function of $\mu$.}
    \label{fig:iso-M_fallback}
\end{figure}

Figure~\ref{fig:iso-M_fallback} shows the simulation results of the tidal disruption of a 1~$M_\odot$ star by a 5, 10, 30, 60 and 90 $M_\odot$ BH  with the default TDE setup in   {\tt Phantom}, which does not include shock heating from stream-stream collisions and radiation pressure.  The penetration factor is assumed to be $\beta=1$ in all the simulations. The mass of the accretion disk, the accreted material, the stream and the remnant core are estimated by the method discussed in Section~\ref{sec:method}.  The upper panel shows the mass of the remnant core (identified as the unbound material) as a function of time.  We can see that, for BHs with mass equal to 10, 30, 60 and 90 $M_\odot$, the mass of the remnant core is around 0.5 $M_\odot$. However, for the 5~$M_\odot$ BH, the disruption is closer to a full TDE, where the mass of the remnant core is $\sim 0.1 M_\odot$. This is also the only case, among the ones we studied, in which the remnant core returns and has a second encounter with the BH. The second encounter happens at $t\simeq 3\times10^5$~s, and produces the bump seen in the green line (upper panel of the figure) around that time.  Since the mass of the remnant core in the $5\,M_\odot$ BH case is very different than the one for the cases with larger BH masses,
there cannot be a uniquely defined variable $\mu$ if we include the $5\,M_\odot$ BH simulation. Hence 
in the following exploration of the dependence of the fallback rate on the mass of the BH we exclude the 5~$M_\odot$ case.

The middle panel of Figure~\ref{fig:iso-M_fallback} shows the fallback rate as a function of time for different BH masses. The SPH simulations by
\citet{Coughlin2019} and \citet{Miles2020} revealed that the fallback rate of the partial TDE scales with $t^{-5/3}$ at early times and switches to $t^{-9/4}$ after a ``break time",
\begin{equation}\label{eq:break}
    t_{\rm break} = \alpha \bigg(\frac{R_\star}{2}\bigg)^{3/2}\frac{2\pi M_\bullet}{M_\star\sqrt{GM_\bullet}}\,,
\end{equation}
where $\alpha$ is a scaling factor dependent on the penetration factor and on the mass ratio of the surviving core to that of the original star, $M_{\rm c}/M_\star$. Generally, $\alpha$ is in the range $\sim 1- 100$. For a partial TDE by a stellar mass BH with $\beta=1$ and $M_{\rm c}/M_\star\sim 0.5$, the typical value of $\alpha$ is $\sim 10$ \citep{Coughlin2015}. Therefore, the typical break time for the partial TDEs in our simulations, as given by Equation~(\ref{eq:break}), is $\sim 10^5$~s. As we can see in the bottom right panel of Figure~\ref{fig:iso-M_fallback}, the late time fallback rate for different BH masses scales with different power-laws, as expected. By fitting the data for times $t>t_{\rm break}$, we find that the late time fallback rate asymptote to -2.10, -1.77, -1.73 and -1.72 for a 10, 30, 60 and 90 $M_\odot$ BH, respectively. Hence we conclude that the powerlaw of the late-time fallback rate of partial TDEs depends significantly on the BH mass of the black hole in the stellar mass BH regime, with a steeper decay  for lighter BHs.

\citet{Coughlin2019} found in their simulations that the powerlaw of the fallback rate of partial TDEs by SMBHs varies from $t^{-5/3}$ to $t^{-9/4}$ as a function of partial fraction (see the right panel of Figure~2 in \citealt{Coughlin2019}). For a full TDE, the late time fallback rate asymptotes to $t^{-5/3}$, while for partial TDEs with varying partial fractions, the early time fallback rate changes from $t^{-5/3}$ to $t^{9/4}$ as the partial fraction increases (larger remnant stellar core). However, for all partial TDEs with different partial fractions, the late-time fallback rate asymptotes to $t^{-9/4}$ so that the late-time fallback rate is practically independent of the mass of the SMBH. However, for partial TDEs by stellar mass BHs, we find that the late-time powerlaw index of the  fallback rate still varies from $t^{-5/3}$ to $t^{-9/4}$ and does not converge to $t^{-9/4}$.

The powerlaw slope of the fallback rate at late times calculated from the fits is different than the one calculated from Equation~(\ref{eq:xi}) that is shown in Figure~\ref{fig:n_infty}. Equation~(\ref{eq:xi}) is a simplified equation that assumes that the late time fallback rate is dominated by gas near the marginally-bound radius. This is a good estimate if the fallback rate is approximated from the accretion rate at the pericenter. However, our new method computes the fallback rate at the separation point between the accretion disk and the fallback stream (as described in Section~2.4 and Figure~\ref{fig:a-dist}), which can be far away from the marginally-bound radius, especially if the accretion disk is inflated by shock heating and radiation pressure. 


Here we provide a fitting formula for the leading order of the late-time  powerlaw index of the fallback rate as a function of $\mu = M_{\rm c}/M_\bullet$ in the stellar mass BH regime. As indicated in the bottom panel of Figure~\ref{fig:iso-M_fallback}, we obtain the fitting formula to be
\begin{equation}
    n_\infty \sim -2.14+4.48\mu^{1/3}-11.82\mu^{2/3}\,.
\end{equation}
\begin{figure}
      \includegraphics[width=.5\textwidth]{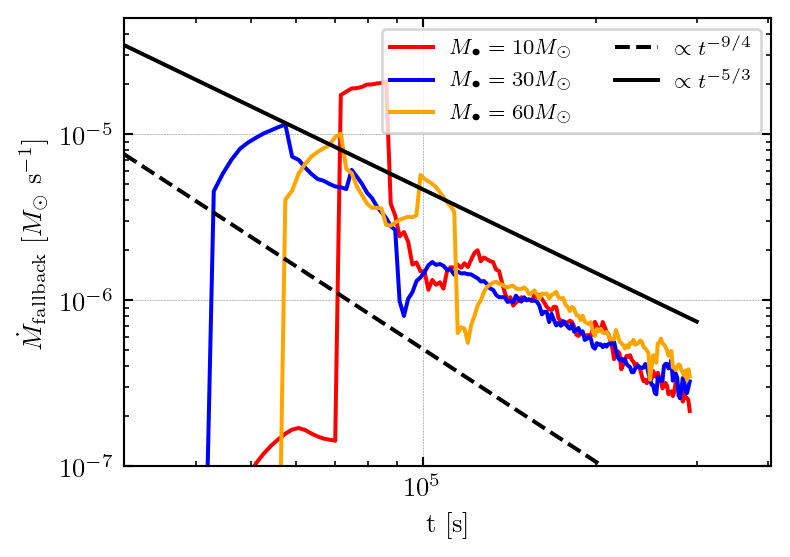}
    \caption{Fallback rate for various BH masses, in simulations which include the effect of radiation pressure. The powerlaw slope at late times for BHs of 10, 30 and 60 $M_\odot$ is -2.18, -1.82 and -1.73, respectively.}
    \label{fig:rad-M_fallback}
\end{figure}

Figure~\ref{fig:rad-M_fallback} shows the fallback rate of TDEs of 1~$M_\odot$ stars by 10, 30 and 60 $M_\odot$ BHs from simulations with radiation pressure included. Fits yield  slopes of the late-time fallback rate of $-2.18$, $-1.82$ and $-1.73$ for 10, 30 and 60 $M_\odot$ BHs, respectively. 
Similarly to the case with adiabatic EoS, we find a dependence of the slope of the 
late-time fallback rate  on the BH mass. However, a more evident stream instability is observed from the simulations. A larger number of collapsed clumps is formed before the stream falls back to the BH.  Although these clumps will be disrupted again when they approach the BH, the fluctuations of the density of the stream caused by these clumps lead to a more variable late-time fallback rate compared to the case without radiation pressure. As we can see from Figure~\ref{fig:rad-M_fallback}, at early times there is a bump in the  fallback rate that is caused by the accretion of the most massive clump.

\section{Summary}
\label{sec:summary}
We have studied the partial tidal disruption of stars by  stellar mass black holes of different mass, using the SPH hydrodynamics code   {\tt Phantom}, in which we implemented an equation of state that includes radiation pressure. The remnant stellar core and the radiation pressure have a significant impact on the instability of the falling back stream and the late-time fallback rate of the debris. Our main results can be summarized as follows:

\begin{itemize}


\item The accretion disks formed from the tidal disruption of stars by stellar mass BHs are optically thick, and the timescale for radiation cooling is relatively long compared to the dynamical timescale. Therefore, shock heating from  stream-stream collisions can significantly heat up the accretion disk.

\item 
Simulations including of shock heating do indeed show that
a larger and thicker accretion disk is formed, rather than the elliptical, thin and confined accretion disk that forms in situations where radiative cooling is efficient. 

\item For TDEs by stellar mass BHs the self-gravity of the falling back stream causes the collapse of the disrupted material into small clumps before accreting onto the BH. These collapsed clumps will cause the fallback rate to fluctuate.

\item In TDEs by stellar mass BHs,  radiation pressure dominates over gas pressure in the inner regions of the accretion disk. As a result, the accretion disk puffs up and some disk material is blown away. The disk is relatively larger (in units of the gravitational radius of the BH). 

\item The late-time fallback rate of  partial TDEs by stellar mass BHs asymptotes to a power-law with different indices that show a strong dependence on the mass of the BH. The power-law  varies from $t^{-5/3}$ to $t^{-9/4}$ as the mass of the BH decreases in the stellar mass regime.

\end{itemize}

\section*{Acknowledgements}

RP acknowledges support by NSF award AST-2006839. PJA acknowledges support from the Simons Foundation under award 644616. The Flatiron Institute is supported by the Simons Foundation.

\section*{Data Availability Statements}
The data underlying this article will be shared on reasonable request to the corresponding author.

\bibliographystyle{mnras}
\bibliography{refs}
\bsp	
\label{lastpage}
\end{document}